\documentstyle[aps,prd]{revtex} 

\draft

\begin{document}
\title{%
\hbox to\hsize{\normalsize\rm March 1999
\hfil MPI-PTh/99-11}
\vskip 36pt
 Decoherence - Fluctuation Relation and Measurement Noise 
}
\author{L.~Stodolsky}
\address{Max-Planck-Institut f\"ur Physik 
(Werner-Heisenberg-Institut),
F\"ohringer Ring 6, 80805 M\"unchen, Germany}
\maketitle
\begin{abstract} We discuss fluctuations in the measurement process
and how these fluctuations are related to the
dissipational parameter characterising quantum damping or
decoherence. On the example of the measuring current of the
variable-barrier or QPC problem we show there is  an extra
noise or fluctuation connected with the possible different outcomes
of a measurement. This noise has an enhanced short time component
which could be interpreted as due to ``telegraph noise'' or
``wavefunction collapses''. Furthermore the parameter giving the 
the strength of this noise is related to the 
parameter giving the rate of damping or decoherence.
 
\end{abstract}
\vskip2.0pc


\section{Introduction}
 I have always shared Lev Okun's interest in the fundamentals of
quantum mechanics, and it is a pleasure to share the following
thoughts with him for his 70th birthday.

Our topic has to do with ``quantum damping'' or ``decoherence'', 
the description of how a quantum system loses its coherence when in
contact with an external system or environment. This is an
interesting and amusing
subject with many aspects. In studying the loss of coherence one
may say we are seeing how a quantum system ``gets classical''. On
the other hand the environment in question may be a ``measuring
apparatus'' and so we get involved with the ``measurement
problem''. Finally, in a more pedestrian vein, many of the problems
and equations are those of ordinary kinetic theory. We have studied
these issues  over the years and applied
the results in many contexts, ranging from optically active
molecules ~\cite{us} to neutrinos~\cite{raf} to gravity~\cite{acta}
and quantum
dots~\cite{qpc}.
In the particularly simple case of the two-state system, such as
the two states of a handed molecule, two mixing neutrino flavors,
or an electron tunneling between two quantum dots, it is possible
to give a fairly  complete phenomenological treatment of the
problem.

Here we would like to discuss a further idea, that there are
certain  fluctuations in the measuring system and these are
connected with the parameter characterizing the damping or
decoherence of the observed system. 
\section{   Damping - Decoherence Parameter}

In the description of the loss of coherence a certain parameter $D$
arises, which can be
thought of as the quantum damping or decoherence rate. We first
describe how this parameter arises. 
 Our description of the two-state system is in terms of the
 density matrix, which is characterized  by a three component
``polarization
vector'' $\bf P$, via 
\begin{equation}\label{rho}
\rho={1\over 2} (I+\bf{P\cdot\sigma})
\end{equation}
where
the  $\bf \sigma$ are 
the pauli matrices.
  $P_z$ gives the
probability for finding  the system in one of the two states
$(\nu_e$ or $\nu_{\mu},$ electron on the left or right dot and so
on) via
$P_z=
Prob(L)-Prob(R)$.  (We shall refer to our two states  as L and R).
Hence $P_z$ gives the amount of the ``quality'' in question. 
 The other components,  $\bf P_{tr}$,  contain 
information on the nature of the coherence. 
 $\vert {\bf P} \vert=1 $ means the system is in a pure state,
 while $\vert {\bf P} \vert=0 $
means the system is completely randomized or ``decohered".  
  $\bf P$ will both rotate in time
due to the real energies in the problem and shrink in length due to
the
damping or decoherence.
 The time development of $\bf P$  is
given by a ``Bloch-like" equation~\cite{us}

\begin{equation}\label{pdot}
{\bf \dot P} = {\bf V x P} - D {\bf P}_{tr} 
\end{equation}

The three real energies $\bf V$ give the evolution of the system in
the absence of damping or decoherence, representing for example the
neutrino mass matrix or the energies for level splitting and
tunneling on the quantum dots.

The second term of Eq.~(\ref{pdot}), our main intereset here,
describes the damping
or
decoherence. $D$
gives the rate  at which
correlations 
are being created between the ``system" (the neutrinos, the
external electron on
the dots)
and the environment or  detector, and this is the rate of damping
or
decoherence.
 The label ``tr"  means ``transverse" to the $z$
axis. The damping only affects ${\bf P}_{tr}$
because 
the ``damping''  or ``observing'' process does not induce   jumps
from one
state to another: the neutrino interaction with the medium
conserves neutrino flavor,  the read-out process for the quantum
dots  leaves the electron being observed on
the same dot; the observing process conserves $P_z=<\sigma_z>$. 

A formula for $D$ can be given~\cite{us} in terms of the S
matrices 
for the interaction of the 
environment or measuring apparatus  with the two states of our
system.
There is a certain complex quantity $\Lambda$ whose imaginary part
gives the
damping $D$ and whose real part gives an  energy shift to the 
system being measured. 
 $\Lambda$ is given by 

\begin{equation}\label{lam}
\Lambda= i (flux)<i\vert{ 1-S_LS_R^\dagger  }\vert i>
\end{equation}

The factor $flux$ is the flux or probing rate of the detector
electrons,
where in the QPC application to be discussed  one
can use the  Landauer
formula $flux = eV_d/\pi\hbar$,
with $V_d$ the voltage in the detector circuit~\cite{thou}). The
label $i$
refers to
the 
the initial or incoming state of the electrons in the detector
and the $S$'s are the S matrices for the two barriers 
corresponding to the 
different states of the observed system.  

 In our   problem with two barriers (L and R) we may express
the barrier penetration problem in S matrix form, calculate
$\Lambda$ according to the above formula and take the imaginary
part to finally obtain $D$~\cite{qpc}. In doing so, we find that
$D$, as might have been expected, involves the difference in
transmission by the two barriers. However, because of the various
phases which are in general present in S, there are some further
and
more subtle phase-dependent effects which can contribute to $D$.
These effects are quite interesting, (and not uncontroversial) but
we shall  not discuss them here and  simply confine ourselves to
the most straightforward situation where
 the only  contributions to $D$ are due to the difference in
transmission by the two barriers. Thus the ``measurement'' consists
solely in the fact that each barrier passes a different current.
With the S matrix parameterized such that the transmission 
coefficients for the two barriers are called $cos\theta_L,
cos\theta_R$, corresponding to transmission probablilities $p_L=
cos^2\theta_L$,$p_R= cos^2\theta_R$,
 we find ~\cite{qpc} with this neglect of phases 
\begin{equation}\label{dama}
 D= (flux)\big\{1- cos \Delta \theta
 \big\}
\end{equation}
where $\Delta \theta =\theta_L -\theta_R$. Hence $D$ is maximal for
very different transmission probabilities or large $\Delta \theta$,
while for
the  case of $\Delta \theta$  small:

\begin{equation}\label{damab}
 D \approx (flux){(\Delta \theta)^2\over 2}
\end{equation}

 $D$ is a phenomenological parameter representing a kind of
dissipation,  resembling in some ways, say, the electrical
resistance. 
Now for resistance and similar dissipative quantities there is the
famous fluctuation-dissipation theorem ~\cite{callen} which
relates the resistance or similar parameter to fluctuations in the
system, as in the relation between resistance and Johnson noise.
Should there be such a relation here? 
 Of course here it is not
energy that is being dissipated. Rather it is ``coherence'' that  
is being lost or perhaps entropy that is being  produced.
Nevertheless we
should expect some such relation~\cite{chandler}. Indeed, looking
at the
Bloch equation Eq~[\ref{pdot}] in its original context as the
description of
the polarization in nuclear magnetic resonance, it is quite natural
to see  the decay of the polarization as due to fluctuating
magnetic fields in the sample.

  Here, however, we wish to consider not fields in a sample but 
something  being observed
by a ``measuring apparatus''.  We will nevertheless reach a similar
conclusion, in that
a ``measuring apparatus'' is something that reacts differently
according to the state of the thing being observed. If it doesn't
react differently, it doesn't ``measure'', obviously. Hence we
expect a measuring apparatus to show fluctuations related not only
to the state of the system being observed, but also to how strongly
it reacts to differences in that system. Furthermore since $D$,
according to
Eq [\ref{dama}],   is
determined by these same differences, we expect some relation
between the damping or decoherence rate and the fluctuations.

  In the following we
would like to show  how such expectations
are realized in the
variable barrier or QPC (quantum point contact) measuring process,
where the ``measuring apparatus'' is a current determined by a
variable tunneling barrier.

\section{The Current in the Variable-Barrier Problem}

Briefly, the measurement process using a  quantum point contact
(QPC)
detector~\cite{buks}
can be described as the modification of a barrier whose
transmission
varies~\cite{field} according to whether an external electron is
nearby 
or farther away.
When the external electron is close by there is a certain higher 
barrier,
and when it 
is farther away, there is a reduced barrier. Given an incident 
or probing flux on the barrier ( in practice also electrons), 
 the modification of the resulting current through the barrier,
 thus ``measures"
where the external electron is located. The state of this external
electron, in the case where it represented as a two-state system is
given by the density matrix, evolving as in Eq[~\ref{pdot}].  The
density matrix
elements 
$\rho_{LL}$ and  $\rho_{RR}$, give the probabilities of the  system
being observed being found in the state L or R
($\rho_{LL}+\rho_{RR} =1$ and $\rho_{LL}-\rho_{RR} =P_z$).

Experiments of this type give a fundamental insight into the nature
of measurement, and 
in an elegant  experiment  Buks et al.~\cite{buks},-
stimulated by 
the work of Gurvitz~\cite{sug}- saw the expected loss of fringe
contrast
in an electron interference arrangement when  one of the
interferometer paths 
 was ``under observation" by a QPC.

Here, however, we wish to focus not so much on the object being
measured but rather on the behaviour of the ``measuring
apparatus''--- the current through the variable barrier or QPC.
 We wish to show is that there are certain ``extra'' fluctuations
in this
current  due to the measurement process.

 To
see this,   we examine the current through the two-barrier system.
 Consider the probability for  a given sequence of transmissions
and reflections through the barriers.
Let  1
represent  a transmission and  0
a reflection for the probing electrons.
Also let $p$ be the probability of transmission and $q$ that for a
reflection, ($q+p=1$), where each of these quantities  has a label
L or R. We write the probability
 that in N probings  
  the first probing electron was transmitted, the second
reflected, ...
 the $N-1$th transmitted and 
the Nth transmitted,  as  $Prob[11...01]$.

 Now it is relatively easy to write down a formula for this
probability in the situation where the time in which the $N$
probings take place is small compared to the time in which  the
observed system changes its state. Taking there to many probings in
this time, we find 
\begin{equation}\label{pro}
Prob[11...01]=\rho_{LL}(p_Lp_L...q_Lp_L)+\rho_{RR}(p_Rp_R...q_Rp_
R),\quad\quad\quad N<<N_{max}
\end{equation}
 where  $N_{max}$ is  the number of probings  in the time it takes
the observed system to change states. The main point about this
formula is that each sequence contains factors of only the $L$ or
$R$ type. We need the restriction $N<N_{max}$ because for longer
times the observed system can change states and the string $
...q_Lp_L$ will get contaminated with factors with $R$ labels. For
short times, where this is not a problem, the
formula applies and may be arrived at either by  thinking about the
amplitude for the whole multi-electron process or by repeated
wavefunction ``collapses''~\cite{qpc}.
 Using it, one can understand the
continuum from ``almost no measurement'' to ``practically reduction
of the wavefunction'' by varying the parameters $p_L$ and $p_R$.
The former occurs
for approximate equality of $p_L$ and $p_R$ and the latter in  the
opposite limit where, say, $p_L$ is
one and $p_R$ is zero.
 One can also understand the origin of ``telegraphic''
signals resembling a ``collapse of the wavefunction'' in this 
latter case: for  $p_L$ and
  $p_R$  close to zero and one respectively,  we will have
predominantly  sequences of either
transmissions or reflections with high probability, while mixed
sequences are
improbable~\cite{qpc}.

Here we would like to use Eq~[\ref{pro}] to examine  the 
fluctuations in the current.

\section {Measurement Noise}
Eq~[\ref{pro}] says that for short times $N<<N_{max}$ we have a
simple combination of two processes,
each one consisting of a sequence characterized by statistical
independence. Naturally a distribution consisting of the
(normalized) sum of two such distinct distributions will have a
variance
greater than  the average of the variances from individual
distributions. For example,  in the classical limit we would have
two distinct
peaks for the transmitted current. Hence we expect  greater
fluctuations than we would have
with just a single distribution.

To quantify this we calculate the variance $V=\overline {Q^2}-
{\overline Q}^2$ in the number of particles transmitted in N
probings of the barrier.
To do this we need
  the probability $Prob(Q,N)$ for Q transmissions in N probings.
 Since 
 the two terms of
Eq~(\ref{pro})  are essentially those leading to 
the binomial distribution in statistics,
 the
combinatorics are that
of the binomial distribution,  and we have
\begin{equation}\label{proba}
Prob(Q,N)=\rho_{LL}{\cal P}_L(Q)+\rho_{RR}{\cal
P}_R(Q),\quad\quad\quad N<<N_{max}
\end{equation}
 where  ${\cal P}_L(Q)$ is the binomial
expression for the probability of Q transmissions in N trials given
the single trial
probability
$p_L$. (For N large and p small this is approximated by the poisson
distribution ${\cal P}_L(Q)\approx {(\bar Q_L)^Q \over Q!} e^{-\bar
Q_L}$, with $\bar Q_L=p_LN$).

Now calulating $V$ for this distribution, that is the variance in
the number of particles
transmitted in N trials,  we obtain

\begin{equation}\label{var}
V(N)=
 \rho_{LL} V_L
+\rho_{RR} V_R+\rho_{LL}(1-\rho_{LL})(\overline Q_L-\overline
Q_R)^2, \quad\quad\quad N<<N_{max}
\end{equation}
where $V_L, V_R$ are the variances for the distributions ${\cal
P}_L,{\cal P}_R$, and we have used $\rho_{RR}=(1-\rho_{LL})$. 

The first two terms are  just  the  average of the variances
of the component distributions, as might have been expected. 
The last term, however, represents some extra noise or fluctuation
due to the fact that a measurement is taking place, that the
current responds differently to the two different states of the
``observed''. 

 A justification for calling  the last term an ``extra'' noise is
also provided when we recognize that the first terms in
Eq [\ref{var}] are  simple
shot noise -like contributions (in ref.\cite{pha} called
``partition
noise'') that would disappear in the current if the
electron charge were not quantized, i.e. in the limit of finite
current but with $e\rightarrow 0$ (see below).

Finally, note that, as is reasonable, this extra noise is absent if
the observed
system is definitely in one state or the other, if either
$\rho_{RR}$ or $(1-\rho_{LL})$ are zero. If only one state is
present there is only one thing to measure and so no extra
fluctuations. Henceforth to simplify the writing  we concentrate on
the relaxed state of the two-state system and put
$\rho_{LL}=\rho_{RR}=(1-\rho_{LL})=1/2$.

 These considerations were all  for $N<<N_{max}$. What happens for
$N>>N_{max}$, for intervals long compared to any transition or
relaxation time? 
Well, then we have $N/N_{max}$ independent, uncorrelated intervals,
each one with a $V$ given by Eq~[\ref{var}]. The variance from  j
independent objects, each one with variance $V_j$ is $V=V_j  j$, so
for long intervals we have from Eq~[\ref{var}] with $N=N_{max}$,
and so
$V_L=p_LN_{max}$ 

\begin{equation}\label{long}
V=V_L +V_R +{(p_L-p_R)^2\over 4} N_{max}N=({p_L+p_R \over 2} 
+{(p_L-p_R)^2\over 4}
 N_{max})N
\end{equation}
going as $N$ or the time, as it should be. 
 
We draw two main conclusions, that there is an extra
fluctuation beyond the simple partition or shot noise-like
contibutions $V_L$,$V_R$
and that this contribution involves the measuring strength or
analyzing power squared  $(p_L-p_R)^2$.

\section{Relation between $D$ and $V_M$}

 Both the extra fluctuations  and  the damping $D$ are
related to the difference in transmission coefficients.
  Calling  this extra ``measurement noise'' $V_M\equiv V-
(\rho_{LL}
V_L+\rho_{RR} V_R)$ we have

\begin{equation}\label{vm}
V_M\sim(\overline Q_L-\overline
Q_R)^2 =(cos^2\theta_L-cos^2\theta_R)^2 N^2,\quad\quad\quad
N<<N_{max} 
\end{equation}  and for 
$D$
 we have Eq~[\ref{dama}]. This relation is implicit but we can find
a simple explicit relation
in the approximation of  working to lowest  order
in $\Delta \theta = \theta_L-\theta_R$:

\begin{equation}\label{vma}
(\overline Q_L-\overline
Q_R)^2\approx 4 p(1-p) N^2 (\Delta \theta)^2
\end{equation}
 where $p$ is the average transmission probability $(p_L+p_R)/2$.
The factor $p(1-p)$ reflects the fact that with all
transmissions or all reflections there are no fluctuations. Now
since   both $V_M$ and $ D$ are proportional to $(\Delta \theta)^2$
we can write a ``decoherence-fluctuation relation'' relating $D$ to
the extra fluctuation $V_m$.

\begin{equation}\label{dv}
V_M\approx ~p(1-p)N^2~(D/flux)\quad\quad\quad
N<<N_{max} 
\end{equation}
 for short times,
while for long times

\begin{equation}\label{dva}
V_M\approx p(1-p)N_{max}N~(D/flux)\quad\quad\quad
N>>N_{max}
\end{equation}

Hence  $D$ can be seen either through the loss of
coherence in the ``observed system'' or through the fluctuations of
the
``measuring system'', here the current.

\section{Conclusions and Experimental Implications}
 With quantum dots and  QPC's  an experimental realization of these
ideas could be one
in which the two-state system controlling  the variable barrier is
an  external electron  resident on and tunneling  between two
dots~\cite{qpc}.
Other variants, both for the two-state system and the probing
current or beam, can undoubtedly be contemplated, for example a  
 single particle or atom in a trap continually probed
by some beam. Note that we are now discussing the totally relaxed
condition of the two-state system, so unlike the situation where we
study oscillations ~\cite{us,qpc} no particular starting time or
injection time need be experimentally defined. We just have certain
extra fluctuations even when  measuring in the incoherent or
relaxed state.  The extra fluctuations  may be said to be due to 
finding sometimes one state and sometimes  the other. 

For  comparison with experiment,
we reexpress the above results in the more usual language of a
current and its variance. It should be noted, however, that we have
not accounted for those fluctuations that may  already
present in the incoming flux, as  from the reservoirs providing the
current~\cite{pha}.

 The parameter N, which  played an important role in the above, was
the number of squential probings under consideration. N is
essentially a measure of the time and may be considered as
giving a  time interval $\Delta t= N/flux $ to be  used in defining
the
current. That is, a certain averaging or integration time 
$\Delta t$ will be involved in converting the number of
transmissions $Q$ into a current $j= e Q/\Delta t$. We thus
introduce
a label $\Delta t$ on the current to indicate the averaging time or
number of probings
used in its definition. This averaging time may be purely
theoretical or represent, for example,  the response time of the
instrumentation. With $e$ the electron charge 

\begin{equation}\label{dt} 
j_{\Delta t}\equiv e Q/\Delta t=eQ(flux/N)
\end{equation}

In the above we considered two regimes, $N<N_{max}$ for times short
compared to the relaxation time for the two-level system
controlling the current, and $N>N_{max}$ for long times compared to
this relaxation time.  
 We can now rewrite Eq~[\ref{var}] as a relation for the variance
of the
current   for short averaging times
\begin{equation}\label{vj} 
\overline{( j^2_{\Delta t})}-(\overline {j_{\Delta t}})^2= \big
({p_L+p_R\over
N}+(p_L-p_R)^2\big)(e~flux)^2,\quad\quad\quad N<<N_{max}
\end{equation}
while for long averaging times Eq~[\ref{long}] becomes
\begin{equation}\label{vja} 
\overline {(j^2_{\Delta t})}-(\overline {j_{\Delta t}})^2=
\big({p_L+p_R\over
N}+{(p_L-p_R)^2\over N} N_{max}\big)(e~flux)^2,\quad\quad\quad
N>>N_{max}
\end{equation}

The fluctuations of the current are seen to depend on the time
scale used in its definition. While the common first term
$((p_L+p_R)/N)(eflux)^2=N(p_L+p_R)/(\Delta t)^2 $ is the usual 
statistical
behavior expected from  the $\surd N$ law for $\surd V$, the
``measurement
induced'' $(p_L-p_R)^2$ term shows quite different behavior in the
two regimes. For long averaging times, Eq~[\ref{vja}], it has the
same $N$ behavior as the first term and so gives an additional
constant level to the partition or shot noise-like contribution.
For short averaging times, Eq~[\ref{vj}], the ``measurement'' term
does not decrease with $N$ at all and for short times but  large
$N$ can completely
dominate the fluctuations.

This increased noise at high frequencies, with no ``ultraviolet
cutoff'' is one of the most interesting points  here. These strong
short time fluctuations are connected with
the ``telegraph noise'' alluded to above. If one wishes,  this can
be
thought of as coming from  ``wavefunction collapses''. Since
these ``collapses'' have no time scale, i.e. are instantaneous,
there is nothing at short
times to smooth or ``soften'' the noise. We stress, however,  that 
this ``collapse'' language is not necessary, everything simply
comes from Eq~[\ref{pro}], which can be obtained in an amplitude
approach~\cite{qpc}, without ``collapses''.

 In both Eq~[\ref{vj}] and Eq~[\ref{vja}] one can see, as mentioned
earlier, that only the
``measurement'' terms  survive in the limit where the charge
$e\rightarrow 0$ and $N\rightarrow \infty$ such that 
the current remains finite. This shows how the noise represented by
the first terms has its origin in the discreteness of the charge,
while that of the second terms has to do with something else,
``measurement''. 

Finally, we rexpress the results of section V  in terms of the
current. Using again the subscript ``M' for the extra,
``measurement induced'' fluctuation we have for  Eq~[\ref{dv}]   
\begin{equation}\label{dvaa} 
\big[\overline{( j^2_{\Delta t})}-(\overline {j_{\Delta
t}})^2\big]_M\approx 8 p(1-p)D (e^2flux),\quad\quad\quad N<<N_{max}
\end{equation}
for short integrating times, with the ``hard'' behavior that it
does not decrease with $\Delta t$, while Eq~[\ref{dva}] becomes

\begin{equation}\label{dvbb} 
\big[\overline{( j^2_{\Delta t})}-(\overline {j_{\Delta
t}})^2\big]_M\approx  8p(1-p){N_{max}\over N}D
(e^2flux),\quad\quad\quad N>>N_{max}
\end{equation}
for long integrating times, with a conventional $1/\Delta t$ fall
off . These express, in the two  regimes, the
``decoherence-fluctuation''relation.

\end{document}